\documentclass[12pt,preprint]{aastex}

\newcommand{\etal}{et al. }

\shorttitle{Evolution and Eddington Ratio Distribution of CT AGN}
\shortauthors{Draper \&  Ballantyne}

\begin{document}

\title{The Evolution and Eddington Ratio Distribution of Compton Thick Active Galactic Nuclei}


\author{A. R. Draper and D. R. Ballantyne}
\affil{Center for Relativistic Astrophysics, School of Physics,
  Georgia Institute of Technology, Atlanta, GA 30332}
\email{aden.draper@physics.gatech.edu}

\begin{abstract}
Previous studies of the active galactic nuclei (AGN) contribution to the cosmic 
X-ray background (CXB) consider only observable parameters 
such as luminosity and absorbing column.  Here, for the first time, we extend the 
study of the CXB to physical parameters including the Eddington ratio of the 
sources and the black hole mass.  In order to calculate the contribution to the 
CXB of AGN accreting at various Eddington ratios, an evolving Eddington ratio 
space density model is calculated.  In particular, Compton thick (CT) AGN are 
modeled as accreting at specific, physically motivated Eddington ratios 
instead of as a simple extension of the Compton thin type 2 AGN population.  
Comparing against the observed CT AGN space densities and $\log N$--$\log S$ relation 
indicates that CT AGN are likely a composite population of AGN made up of sources
accreting either at $>$ 90$\%$ or $<$ 1$\%$ of their Eddington rate.
\end{abstract}


\section{Introduction}
\label{sect:intro}

It is believed that all forms of active galactic nuclei (AGN) are variations 
of the same phenomenon, namely accreting supermassive black holes \citep{R84}.  
The most basic observational description of AGN is their bolometric luminosity, 
$L_{bol}$, and the obscuring column density, $N_{\mathrm{H}}$, through which they are observed.  
AGN are known to radiate at luminosities $L_{bol}\approx$ 10$^{40}$--10$^{48}$ 
erg s$^{-1}$, but this is more physically described by the Eddington ratio, defined 
as $L_{bol}/L_{Edd}$, where $L_{Edd}\equiv 4\pi GM_{\bullet}m_pc/\sigma_T$ is the 
Eddington luminosity of a black hole with mass $M_{\bullet}$.  Observed Eddington ratios range 
roughly ten orders of magnitude with a maximum of $\sim$1 \citep{CX07}.  There 
is a similarly wide range of observed column densities ($20\leq\log N_{\mathrm{H}}\la25$; e.g., Tueller \etal 2008).

The most widely accepted theory to explain the observed range of obscuration is known as the 
unified model and claims that the obscuration of an AGN is dependent solely on the orientation 
of the AGN with respect to the line of sight to the observer (e.g., Antonucci 1993).  However, there is evidence 
that suggests that Compton thick (CT) obscuration, when $\log N_{\mathrm{H}}\gtrsim 24$,
is linked to an evolutionary phase during which the 
black hole and the host stellar spheroid accrete rapidly 
(e.g., Fabian 1999; Page \etal 2004; Rigopoulou \etal 2009).  As the black hole grows, the outflow 
from the black hole strengthens and ejects the obscuring material, revealing an unobscured quasar.  This 
outflow also halts the growth of the host spheroid causing the observed proportionality 
between black hole mass and host spheroid mass (e.g., Crenshaw \etal 2003; Page \etal 2004; Alexander \etal 2010).  Galaxy merger simulations 
show that mergers between gas rich galaxies will cause gas to flow into the nuclear region, 
igniting a nuclear starburst and highly obscured, high Eddington ratio quasar activity (e.g., 
Hopkins et al. 2006).  \citet{Fab09} also show that if an AGN is accreting at close to Eddington, 
radiation pressure on dust will blow out any column density with $\log N_{\mathrm{H}}\la 24$.  
Therefore, if an AGN is obscured and has a high Eddington ratio, it must be CT.  Observational 
evidence also supports this evolutionary scenario.  \citet{P04} found that star formation in 
the hosts of unobscured quasars has already peaked, whereas the hosts of obscured quasars are 
still undergoing massive amounts of star formation indicating that their galactic spheroids are 
still forming.  \citet{T09a} found that the space density of luminous CT AGN candidates evolves strongly at 
$z=$ 1.5--2.5, so that the CT quasar population seems to peak at a slightly higher redshift than the population 
of unobscured quasars.  This evolutionary theory is difficult to observationally test since the high 
levels of obscuration found in CT AGN make them nearly invisible below 10 keV in their rest-frame.

If being CT is an early evolutionary phase of powerful quasars, CT AGN should be very rare at $z\sim0$; 
however, {\em Swift}/BAT and {\em INTEGRAL} have detected 
several CT AGN at $z<0.025$ (Tueller \etal 2008; Malizia \etal 2010).  According to black hole 
masses and X-ray luminosities reported by \citet{Gult09}, the two brightest CT AGN, the Circinus 
Galaxy and NGC 4945, have $\lambda=-2.0$ and $\lambda=-5.6$, respectively, where $\lambda\equiv\log L_{bol}/L_{Edd}$. 
Therefore the scenario where CT AGN are high Eddington ratio quasars does not completely describe the 
entire CT AGN population.  

As the majority of the cosmic X-ray background (CXB) $\la$ 10 keV has been resolved into AGN 
by deep observations by {\em ROSAT}, {\em Chandra}, and XMM-{\em Newton} 
(see Brandt \& Hasinger 2005), the CXB provides a complete census of AGN.  Due to the observational 
constraints of highly obscured sources, CT AGN are difficult to study even in the local universe; thus, historically CT AGN have 
been invoked to augment the observed AGN population in population synthesis models to make these 
models properly fit the observed, but not resolved, peak of the CXB around 30 keV.  As there is a dearth of 
observational information about CT AGN, previous population synthesis models have 
treated CT AGN as a simple extension of type 2 AGN (e.g., Ueda \etal 2003; Treister \& Urry 2005; Ballantyne \etal 2006; Gilli \etal 2007); 
however, if CT AGN are part of an evolutionary sequence, 
then they would not evolve in the same manner as the less extremely obscured type 2 AGN.  Indeed, fitting the CXB 
in that manner seems to overpredict the local observed space density of CT AGN (see Treister et al. 2009a).  

In this letter, we compute the contributions of AGN with various 
Eddington ratios to the CXB and model CT AGN using a physically motivated Eddington ratio distribution.  
We then compare our model predictions with observed cumulative and decadal space densities and the 
{\em Swift}/BAT CT $\log N$--$\log S$ relation.  We find that CT AGN are a composite population of 
AGN accreting at both very high Eddington ratios and very low Eddington ratios.  When necessary, a 
$\Lambda$CDM cosmology is assumed with $h_0=0.7$, $\Omega_M=0.3$, and $\Omega_{\Lambda}=0.7$.

\section{Calculations}
\label{sect:calc}

\subsection{Eddington Ratio Distribution}
\label{sub:erd}

Following the method of \citet{Mer09} and \citet{MH08}, with minor modifications described 
below, we calculate the Eddington ratio space density, 
$\Phi_{\lambda}\left(L_X,M_{\bullet},z\right)$, in Mpc$^{-3}$, where $L_X$ is the 2--10 keV luminosity.  
The bolometric correction from \citet{Mar04}, 
\begin{equation}
\log(L_{bol}/L_X)=1.54+0.24\mathcal{L}+0.012\mathcal{L}^2-0.0015\mathcal{L}^3, 
\label{eq:bolcor}
\end{equation}
where $\mathcal{L}=\log (L_{bol}/L_{\bigodot})-12$, 
is used to transform X-ray luminosities to bolometric luminosities.  \citet{Mar04} derive this bolometric 
correction using an average intrinsic spectral energy distribution template for radio 
quiet AGN. The scatter in the bolometric correction is $\sim$0.1 dex, which is negligible in our analysis.  The black hole mass function (BHMF) is derived from a Gaussian curve fit to the combined type 1 and type 
2 AGN BHMF observed by \citet{Net09} at $z\sim 0.15$.  As the \citet{Net09} BHMF is fractional, the \citet{Mer09} and \citet{MH08} method was modified to use only the space density from the \citet{U03} hard X-ray luminosity function (HXLF), instead of depending equally on the space density from the BHMF and the HXLF.  \citet{MH08} argue that black hole mergers will only have a perturbative effect on black hole growth as the uncertainties in the AGN bolometric luminosity function are larger than the impact of black hole mergers on the local BHMF.  Therefore, the evolution of the BHMF must satisfy the conservation equation (Small \& Blandford 1992):
\begin{equation}
\frac{\partial n_M(M_{\bullet},t)}{\partial t}+\frac{\partial[n_M(M_{\bullet},t)\langle\dot{M}(M_{\bullet},t)\rangle]}{\partial M}=0,
\label{eq:coneq}
\end{equation}
where $n_M$ is the BHMF and $\langle\dot{M}(M_{\bullet},t)\rangle$ is the average accretion rate at time $t$.   
Thus the BHMF is evolved by integrating this conservation equation forward and backward in time, using the observed 
BHMF at z~0.15 as the boundary condition.  The critical 
Eddington ratio where the transition from radiatively efficient thin disk accretion 
to radiatively inefficient geometrically thick accretion is taken as $\lambda_{cr}=-2.0$ 
\citep{CX07}.  For $\lambda>\lambda_{cr}$ the radiative efficiency is assumed to be $\epsilon_{rad}=0.1$ and for 
$\lambda<\lambda_{cr}$, $\epsilon_{rad}=0.1(\lambda/\lambda_{cr})^{1/2}$ \citep{MH08}.

\subsection{Cosmic X-ray Background Model}
\label{sub:CXB}

The CXB model closely follows that described by 
\citet{DB09} with differences in the spectra of CT sources and low Eddington 
ratio sources, and $\Phi_{\lambda}(L_X, M, z)$ is input in place of a luminosity function.  
A torus reflection component, calculated using "reflion" \citep{RF05}, is added to the CT AGN spectrum (cf., Gilli \etal 2007). 
For AGN with $\lambda<-2.0$, the reflection component is included only if the source is CT.  
In order to properly fit the CXB at higher energies, the power-law component of the 
$\lambda<-2.0$ spectrum is assumed to have a cut-off energy of 200 keV, otherwise a cut-off 
energy of 250 keV is used\footnote{If 250 keV is used as the cut-off energy for all sources the qualitative findings of this study do not change, but the fraction of CT AGN is reduced by $\sim$1$\%$.}.  Most importantly, CT AGN are added at specific Eddington 
ratios which are chosen based on a physically motivated scenario instead of adjusting 
the normalization of the luminosity function to account for CT AGN as done with previous 
CXB models.

\section{Results}
\label{sect:res}

\begin{figure*}
\begin{center}
\includegraphics[angle=0,width=0.95\textwidth]{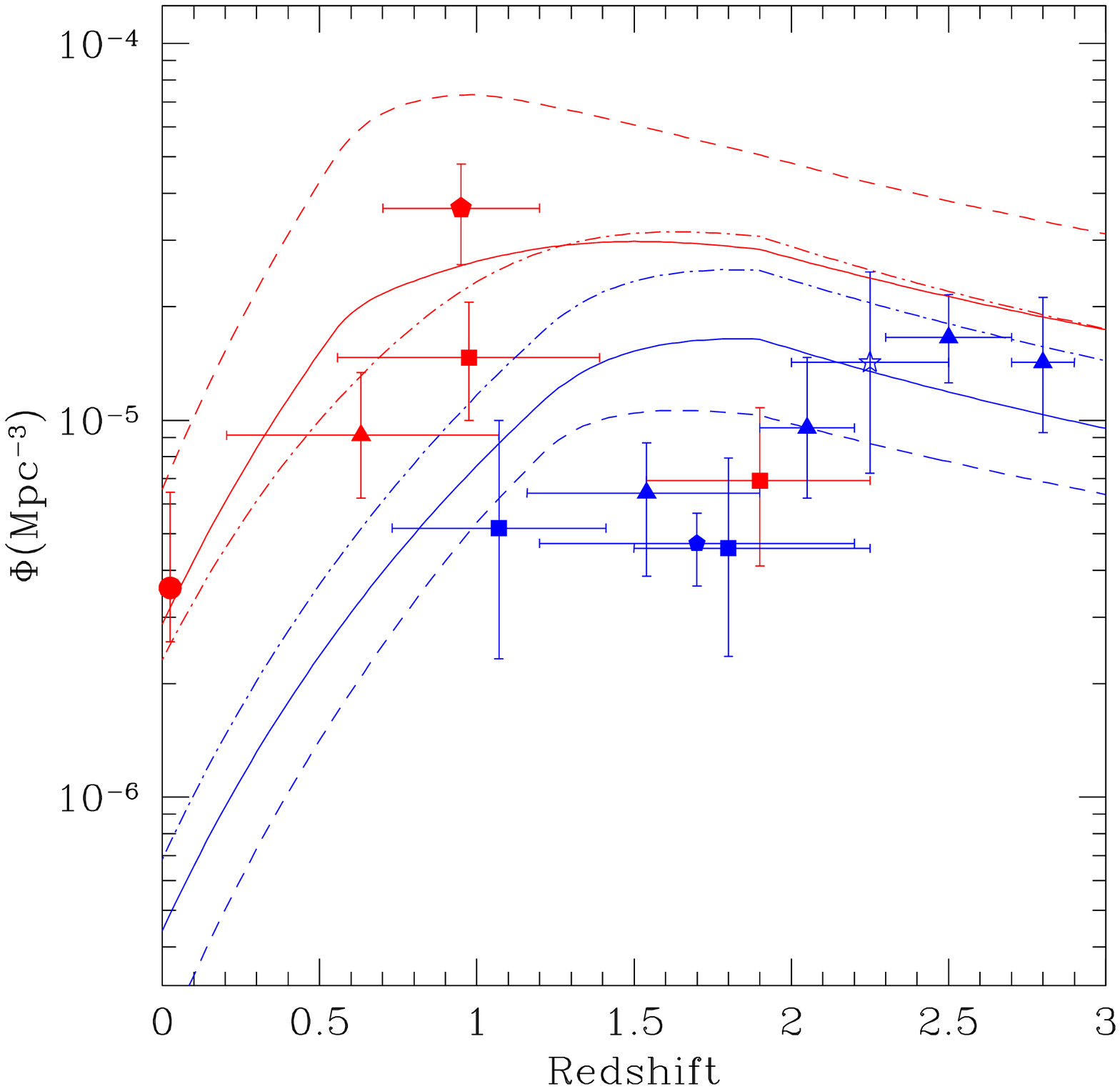}
\end{center}
\caption{Cumulative CT AGN space density.  Red points and lines denote CT AGN with $\log L_X > 43$ and blue points and lines denote CT AGN with $\log L_X > 44$.  Solid lines show the composite model where CT AGN have either $\lambda<-2.0$ or $\lambda>-0.05$. The dashed lines show the original model where CT AGN are considered a simple extension of the Compton thin type 2 population.  The dot-dashed lines show the scenario where all CT AGN have $\lambda>-0.05$.  Data points are shown from several studies: triangles: Treister et al. (2009a); squares: Tozzi et al. (2006); stars: Alexander et al. (2008); pentagons: Fiore et al. (2009); circle: local {\em Swift}/BAT and INTEGRAL data point reported by \citet{T09b} adjusted to reflect the flux-luminosity relation for CT AGN described by \citet{R09}.}
\label{fig:cum}
\end{figure*}

\begin{figure*}
\begin{center}
\includegraphics[angle=0,width=0.95\textwidth]{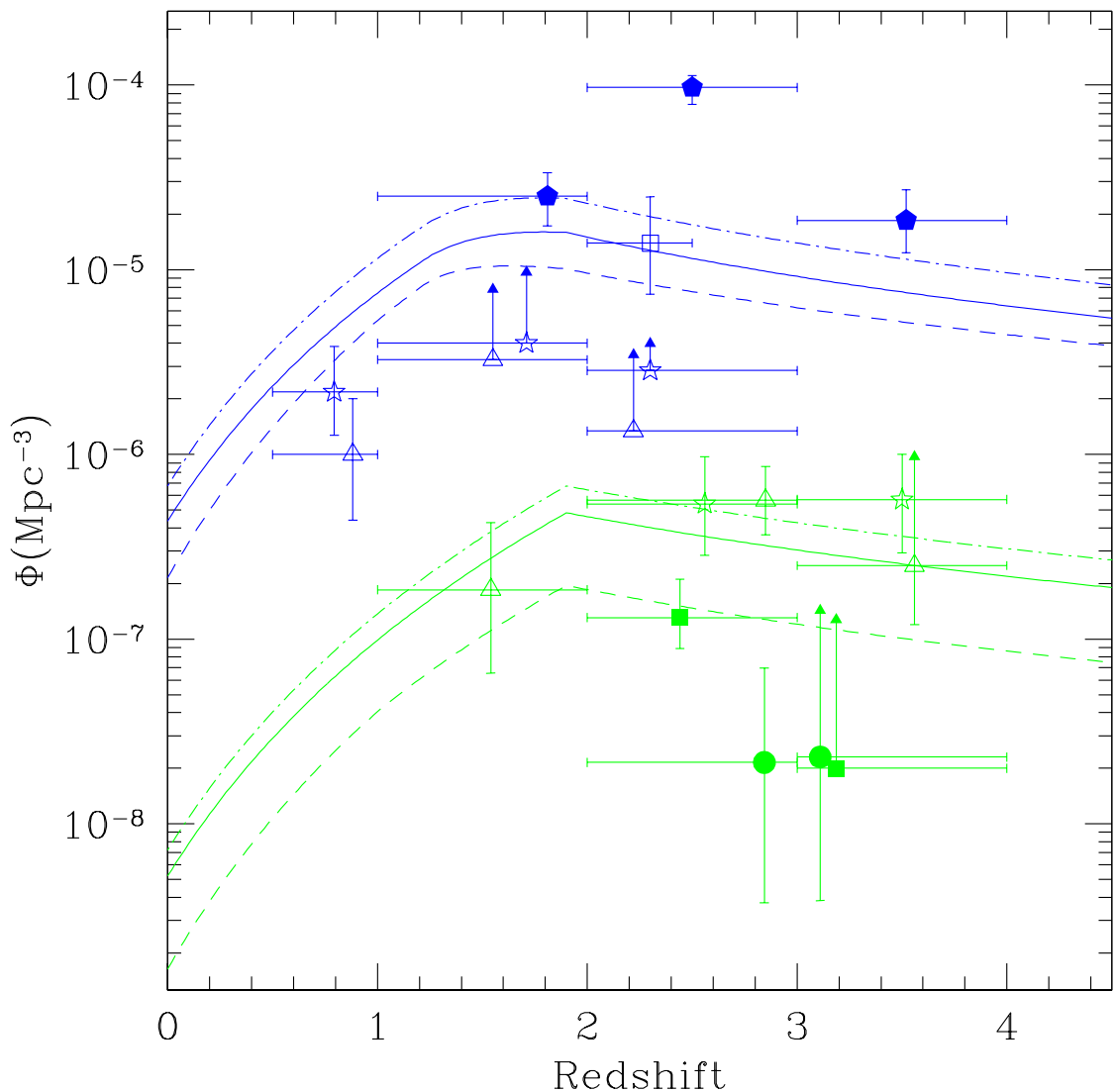}
\end{center}
\caption{Decadal CT AGN space density.  Blue points and lines denote CT AGN with $\log L_X=$ 44--45 and green points and lines denote CT AGN with $\log L_X=$ 45--46.  Lines are the same style as in figure \ref{fig:cum}.  Data points are shown from several studies: filled squares: Yan et al. (2007); circles: Mart\'inez-Sansigre et al. (2006); stars: Polletta et al. (2006); triangles: Fiore et al. (2009); pentagons: Fiore et al. (2008); open squares: Alexander et al. (2008).}
\label{fig:dec}
\end{figure*}

\begin{figure*}
\begin{center}
\includegraphics[angle=0,width=0.95\textwidth]{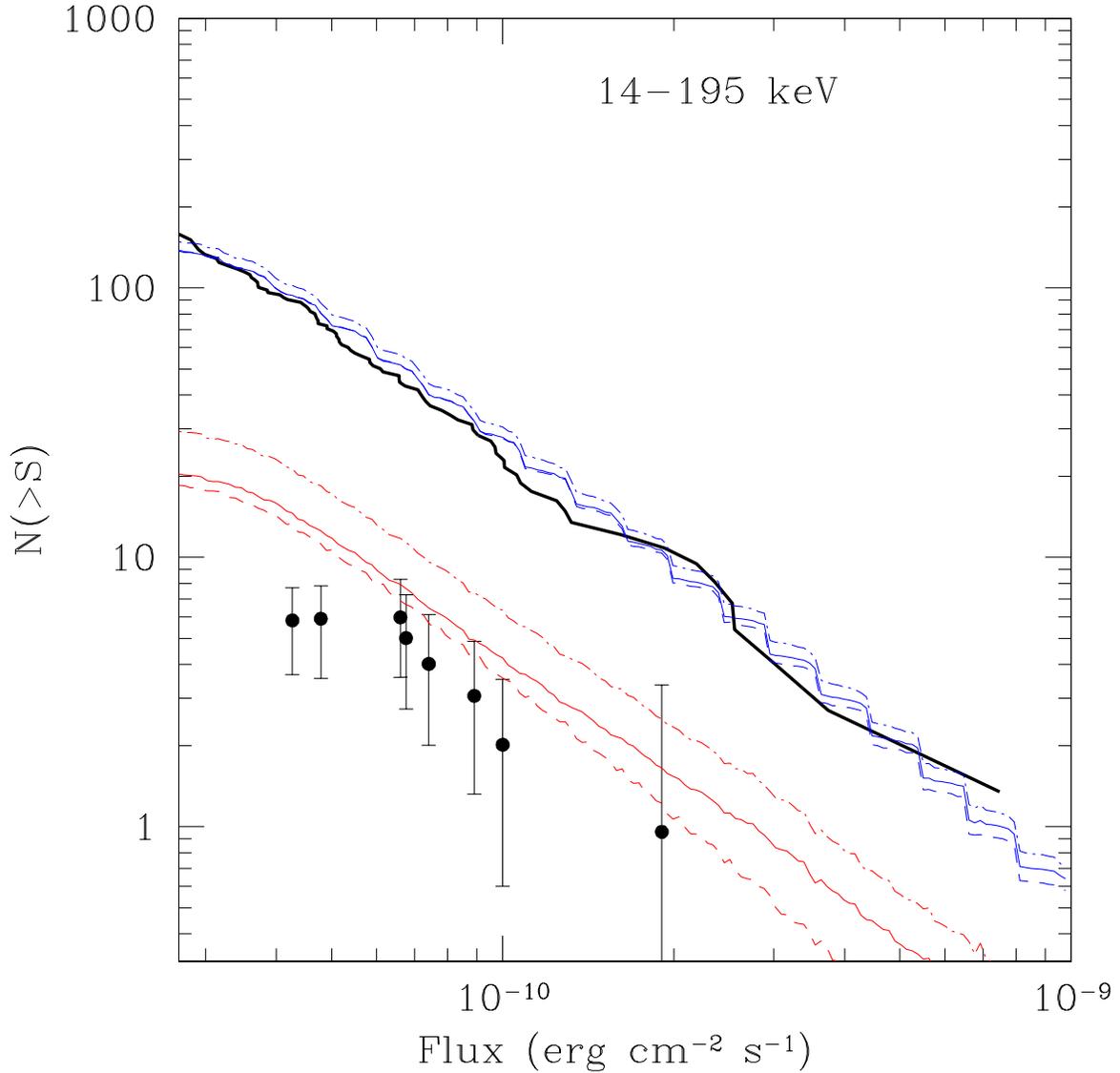}
\end{center}
\caption{BAT number counts.  Blue lines signify total AGN number counts and red lines signify transmission dominated CT AGN number counts.  Colored lines are the same style as in figure \ref{fig:cum}.  The black line shows the total AGN $\log N$--$\log S$ relation from Tueller et al. (2008) and the black data points show the CT $\log N$--$\log S$ data points from Treister et al. (2009b).}
\label{fig:bat}
\end{figure*}

In previous population synthesis models the number of CT AGN was assumed to be proportional 
to the number of type 2 AGN.  In this original model we find 44$\%$ of 
all type 2 AGN are CT, where the normalization is fixed by fitting the peak of the CXB ($EF_E\approx$ 
44.2 keV cm$^{-2}$ s$^{-1}$ str$^{-1}$ at 30 keV).  
In figure \ref{fig:cum} the cumulative space density is shown for CT AGN with $\log L_X>43$ 
(red) and $\log L_X>44$ (blue).  The dashed line shows the predictions of the original 
model.  The red circle shows the local {\em Swift}/BAT and INTEGRAL data originally reported by \citet{T09b} 
and here has been corrected to account for the flux-luminosity relation for CT AGN 
described by \citet{R09}.  The original model significantly overpredicts the 
observed space density of CT AGN with $\log L_X>43$ at all redshifts.  The original model also 
overpredicts the observed space density of CT AGN with $\log L_X>44$ 
at $1.5<z<2$, while underpredicting the $z>2.5$ data points.
The decadal CT AGN space density is shown in figure \ref{fig:dec} where data and predictions for CT AGN with $\log L_X$ = 44--45  
are colored blue, and data and predictions for CT AGN with $\log L_X$ = 45--46 are colored green.  
The dashed lines again show the predictions of the original model.  This model agrees fairly well with 
the $\log L_X$ = 44--45 and $\log L_X$ = 45--46 data at low redshift, but underpredicts the density of high redshift sources.  
As seen in figure \ref{fig:bat}, the original model agrees fairly well with the {\em Swift}/BAT $\log N$--$\log S$ 
relation, where red lines denote the transmission dominated CT AGN number 
counts and dashed lines denote the original model.  This model overpredicts the 
BAT number counts at the lowest flux levels but is in decent agreement with 
observations at the higher flux levels.
A $\chi^2$ test was performed using the cumulative CT space density 
data points, the decadal CT space density data points which are not lower limits, 
and the BAT CT AGN number counts data points, giving 32 degrees of freedom.  
The original model was found to have a reduced $\chi^2$ of $\chi^2_{red}=15$.

We then attempted to model the CT population as only including sources with $\lambda>-0.05$ independent of the fraction of type 2 AGN at any $L_X$ and $z$.  In order to match the peak of the CXB, $\sim$91$\%$ of these AGN at any time must be CT.  As seen in figures \ref{fig:cum}, 
\ref{fig:dec}, and \ref{fig:bat}, where the predictions from this model are shown as the dot-dashed 
lines, this scenario underpredicted the observed local CT AGN space densities, 
overpredicted the {\em Swift}/BAT $\log N$--$\log S$ relation by a factor of 2.0 to 3.5, and resulted in $\chi^2_{red}=25$.  
Thus, the CT population cannot be modeled as only including sources with very high Eddington ratios.

\begin{figure*}
\begin{center}
\includegraphics[angle=0,width=0.95\textwidth]{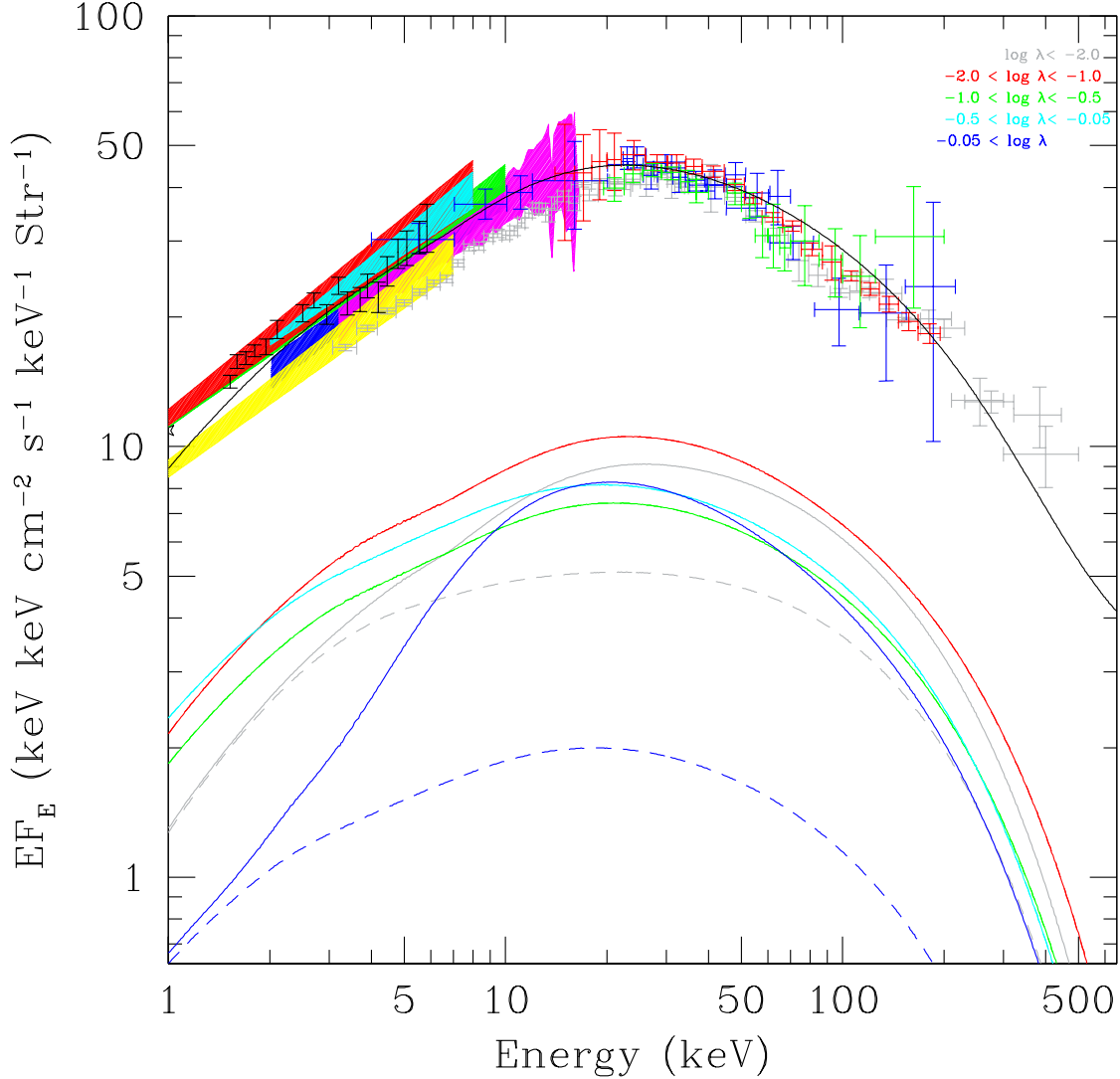}
\end{center}
\caption{CXB fit using the composite model showing contributions of AGN accreting at different Eddington ratios.  The solid black line shows the combined contribution of non-blazar AGN and blazars to the CXB.  Colored lines show the contribution of AGN accreting at the color designated Eddington ratio, where dashed lines refer only to the Compton thin contribution and solid lines show the combined contribution of Compton thin AGN and CT AGN.  Measurements from various instruments are shown as colored areas and data points; blue: {\em ASCA} GIS (Kushino et al. 2002); magenta: {\em Rossi X-ray Timing Explorer} ({\em RXTE}; Revnivtsev et al. 2003); green: XMM-{\em Newton} (Lumb et al. 2002); red: {\em BeppoSAX} (Vecchi et al. 1999); yellow: {\em ASCA SIS} (Gendreau et al. 1995); cyan: XMM-{\em Newton} (De Luca \& Molendi 2004); grey data: {\em HEAO}-1 (Gruber et al. 1999); blue data: {\em INTEGRAL} (Churazov et al. 2007); red data: {\em Swift}/BAT (Ajello et al. 2008); black data: {\em Swift}/XRT (Moretti \etal 2009); green data: {\em INTEGRAL} (T\"urler et al. 2010).}
\label{fig:edd}
\end{figure*}

The scenario where all CT AGN have $\lambda>-0.05$ failed to agree with local observations, and 
as local AGN are known to have low Eddington ratios \citep{H08}, low Eddington ratio sources were 
then added to the CT AGN population model.  In order to fit the peak of the CXB, while not overestimating 
the CXB in the 100--200 keV range, this composite model predicts $\sim$60$\%$ 
of AGN with $\lambda< -2.0$ and $\sim$86$\%$ of AGN with $\lambda>-0.05$, at any time and independent of 
the fraction of type 2 AGN, are CT. The CXB fit found using this 
composite model is shown in figure \ref{fig:edd} with the 
contribution made by AGN accreting at different Eddington ratios plotted as separate lines.  In this model the CXB is still dominated by obscured 
sources accreting at moderate Eddington ratios, i.e. type 2 Seyferts.  The composite 
model predictions are shown as the colored solid lines in figures \ref{fig:cum}, \ref{fig:dec}, and 
\ref{fig:bat}.  This model is in fairly good agreement with the observed space densities 
for CT AGN with $\log L_X>43$ and $\log L_X>44$, except for overpredicting the observed space density of 
CT AGN with $\log L_X>44$ at $1.5<z<2$.  The composite model predictions match the observed cumulative space 
densities better than the original model predictions.  At $z\la 1$, the composite model slightly 
overpredicts the space density of CT AGN with $\log L_X$ = 44--45 and underpredicts the observed 
densities at $z\gtrsim 2.5$.  For CT AGN with $\log L_X$ = 45--46, this scenario agrees 
fairly well with observations.  In general, the composite model is a better fit to the high luminosity, 
high redshift data than the original model.  This scenario does
overpredict the BAT number counts at the lowest flux levels, but is in decent agreement with 
observations at the higher flux levels.  We find $\chi^2_{red}=9.8$, 
indicating that the composite model is in better overall agreement 
with the observed CT AGN space densities and the {\em Swift}/BAT $\log N$--$\log S$ 
relation when compared to the original model, or one where all CT AGN have $\lambda>-0.05$.

Other models were investigated in which the CT population was assumed to be accreting at only 
low or moderate Eddington ratios.  When the CT population was assumed to all be accreting at 
$\lambda<-2.0$ the model underpredicted the 
observed high luminosity CT space densities and slightly underpredicted the {\em Swift}/BAT 
$\log N$--$\log S$ relation.  When $-2.0<\lambda<-0.5$ accretion was assumed 
for the CT population the model overpredicted the observed local space 
density by a factor of 3.2 and significantly overpredicted the {\em Swift}/BAT $\log N$--$\log S$ 
relation.  In conclusion, it is necessary to include both sources with $\lambda<-2.0$ and $\lambda>-0.05$ to model 
the CT AGN population.

\section{Discussion and Summary}
\label{sect:sum}

We find that current data suggests that the majority of AGN accreting at close to the Eddington limit are CT,   
in agreement with galaxy merger simulations (e.g., Hopkins \etal 2006) that show that 
mergers funnel gas into the central regions of galaxies igniting both starbursts and 
high Eddington ratio accretion onto the central black hole.  These AGN are able to accrete rapidly 
because of a large abundance of gas and dust in the central region of the host bulge.  
This abundance of gas and dust naturally leads to a very 
high obscuring column density \citep{F99}.  Furthermore, radiation pressure from a rapidly 
accreting AGN will blow out any dusty gas with $\log N_{\mathrm{H}}\la 24$ \citep{Fab09}.  
Thus, in accordance with our findings, the vast majority of rapidly accreting, obscured AGN must be CT.  

A new result is that the CT AGN population, in addition to including sources with $\lambda>-0.05$, 
also includes sources with $\lambda\la-2.0$.  
While \citet{H08} explains that most observed low Eddington ratio sources tend to be unobscured, \citet{TW03} 
find evidence for a population of highly obscured low Eddington ratio AGN.  Indeed \citet{M10} find an 
obscuration-luminosity relation in their sample of {\em INTEGRAL} sources which claims that lower 
luminosity sources tend to be more obscured.  As these observed lower luminosity sources are local, 
these AGN are likely to have low Eddington ratios.  Since low Eddington ratio AGN are only weakly accreting, 
they will have very little effect on their environment.  Therefore molecular clouds would be able to come 
deep into the core of the bulge without being affected.  Indeed \citet{TW03} find that the highly obscured low 
Eddington ratio sources in their sample might exhibit variable obscuration, as would be expected if the 
obscuration were due to molecular clouds near the black hole. Several studies have been conducted on so called 
'changing-look' AGN (see Bianchi \etal 2005 and references therein), which are AGN whose spectrum changes from 
Compton thin to reflection dominated on the time scale of years.  It is believed that the changes in the spectrum of these 
local, low luminosity AGN is due to either the nucleus being obstructed by a cloud with a CT column 
density or the nucleus switching off, so that the reflection dominated state is actually the echo of a 
previous accretion episode.  \citet{B05} find that it is most likely that the four AGN in their sample are in 
CT states.  In a study of 82 LINERs, \citet{G09} find about half of the LINERs in their sample appear to be CT and that 
there is a higher percentage of CT LINERs than CT Seyferts.  The average Eddington ratio of the CT LINERs in 
the sample used by \citet{G09} is $\lambda=$ -4.7.  This finding is consistent with the composite model for 
CT AGN discussed here which finds that $\sim$60$\%$ of AGN with $\lambda < -2.0$ are CT.

It is clear that previous attempts to model the elusive CT AGN population as an extension of the Compton 
thin type 2 population fail to explain the observed CT AGN space density and $\log N$--$\log S$ 
relation.  \citet{T09b} suggest that CT AGN only contribute $\sim$10$\%$ of the CXB and that the normalization 
of the CXB is overestimated by population synthesis models which predict a large population of CT AGN.  
However, changing the normalization of the CXB does not account for the strong evolution of CT AGN at 
$z=$ 1.5--2.5 found by \citet{T09a}.  Here we assume that CT AGN contribute $\sim$20$\%$ of the 
CXB based on the findings of \citet{DB09}, but the composite model presented here does account 
for the strong evolution of CT AGN at high redshift while not overestimating the local CT AGN space density.  Additionally, 
models which include CT sources with $-2.0<\lambda<-0.5$ overpredict 
locally observed CT AGN space density.  Models which only include CT sources with $\lambda<-2.0$ 
underpredict the density of high luminosity CT sources.  If CT sources are assumed to only have 
$\lambda>-0.5$, the model overpredicts the observed $\log N$--$\log S$ relation.  A composite 
model, where the CT AGN population includes both sources accreting at close to Eddington and 
sources accreting at $<$1$\%$ of Eddington, best explains observations.  

As CT AGN are rare and difficult to observe due to high obscuration, it is 
difficult to use observations to directly decipher the physical phenomenon which gives rise to 
the extreme levels of obscuration found in CT AGN (e.g., Triester et al. 2009b; Rigby et al. 2009).  
Thus physical models of CT AGN must use indirect observational constraints, like observed space 
densities and number counts, to uncover the nature of the extreme obscuration of these sources.  
This letter presents a model which constrains the physical parameter of the Eddington ratios of 
CT AGN to $\gtrsim$90$\%$ and $\la$1$\%$.  In the future the understanding of CT AGN will be expanded 
through IR and sub-mm studies of the reprocessed radiation from CT AGN and very hard 
X-ray imaging with missions like {\em NuSTAR}.

\acknowledgments
The authors thank A. Merloni for helpful conversations.


{}


%
%
%
%

\end{document}